\documentclass[twocolumn, pra,showpacs,superscriptaddress]{revtex4}
\usepackage{amssymb}
\usepackage{mathrsfs}
\usepackage{graphicx}
\usepackage{subfigure}
\usepackage{float}
\usepackage[english]{babel}
\usepackage[utf8x]{inputenc}
\usepackage[T1]{fontenc}
\usepackage{color}
\usepackage{CJK}
\usepackage{natbib}
\usepackage{dcolumn}
\usepackage{bm}
\usepackage{amsmath}
\newlength{\figurewidth}
\begin{document}
\begin{CJK*}{UTF8}{}

\title{Ground state of Tonks-Girardeau gas under density-dependent gauge potential in a one dimensional harmonic potential}
\author{Yajiang Hao \CJKfamily{gbsn}(郝亚江)}
\email{haoyj@ustb.edu.cn}
\affiliation{Institute of Theoretical Physics and Department of Physics, University of Science and Technology Beijing, Beijing 100083, P.~R.~China}

\date{\today}

\begin{abstract}
In the present paper we investigate the ground state of Tonks-Girardeau gas under density-dependent gauge potential. With Bose-Fermi mapping method we obtain the exact ground state wavefunction for the system confined in a harmonic potential. Based on the ground state wavefunction, the reduced one body density matrix (ROBDM), natural orbitals and their occupations, and the momentum distributions are obtained. Compared with the case without gauge potential, the present wavefunction and ROBDM have additional phase factors induced by gauge potential. The momentum distribution is the convolution of that without gauge potential to the Fourier transformation of definite integral of gauge potential. It is shown that because of the density-dependent gauge potential the peak of momentum distributions deviate from zero momentum and the Bose gas take finite total momentum. In particular the momentum distribution is no longer symmetric although the total momentum can become zero by adding a constant to the gauge potential.
\end{abstract}

\maketitle
\end{CJK*}

\section{Introduction}

The interaction between particles plays an important role in the quantum many-body system. For ultracold atoms the strength of atomic interaction can be tuned in the full interaction regime from infinite attraction to infinite repulsion with the Feschbach resonance and confined induced resonance technique by controlling the magnetic field \cite{FR,CIR}. Besides interaction, the dimension effects were also the focus of the study of ultracold atoms to simulate the traditional condensed matter system. Since the experiment realization of Bose-Einstein condensates (BECs) the low dimensional quantum gas has attracted the interests of many experimentalists and theorists. With the highly anisotropic trap or two dimensional optical lattices the ultracold atom can be tightly confined in transverse directions and weakly confined in the axial direction \cite{Paredes,Toshiya,Ketterle,Single1D}. In this way the trapped ultracold atoms become a one-dimensional (1D) quantum gas. The interplay of atomic interaction and dimension significantly stimulated the interest in strongly correlation effect in the low dimensional quantum gas \cite{RMP2011,RMP2012,RMP2013}. It has been shown that with the increase of interaction the ground state density profiles of 1D quantum gas evolve from the Gaussian-like Bose distribution to the fermionized shell structure \cite{Hao2006}.

For the highly controllability of ultracold atoms and the utilization of optical lattices, ultracold atoms has been used to engineer condensed-matter models \cite{RMP2011}. In addition, ultracold atoms can also be used to simulate charged particles in electromagnetism field although only neutral atoms exist in experiment \cite{ RMPDalibard,RPPGoldman}. By coupling different internal spin state \cite{YJLinPRL2009,YJLinNat2009,YJLinNatPhys2011} or lattice sites \cite{Miyake,IBlochPRL2013} with Raman light, or by shaking lattice \cite{Struck,Parker}, the synthetic gauge fields are induced and the spin-orbital coupling BECs can also be produced \cite{SOCBECs}. Not only the quantum gas in static synthetic gauge fields has been investigated, but also the feedback of the atoms on synthetic gauge field attract more and more interests \cite{ChiralSoliton,LSantosPRL2014,LSantosPRB2015,CChinPRL2018,Dingwall}. This is important to simulate the physics related with interaction between gauge field and matter in various fields. It was found that density-dependent synthetic gauge field result in chiral solitons and chiral superfluid \cite{ChiralSoliton,LSantosPRB2015,Dingwall}. It is also expected that density-dependent synthetic gauge field will induce new phase transition \cite{LSantosPRL2014} and even the same techniques and experiment schemes can be used to create 1D anyonic system satisfying fractional statistics \cite{CChinPRL2018,AnyonSantos}.

Although ultracold atoms in synthetic gauge fields and the BECs with spin-orbital coupling have been investigated very much, most of them focus on the weakly interacting BECs. In the present paper we will investigate the ground state of the strongly interacting Tonks-Girardeau (TG) gas in a harmonic potential under synthetic gauge potential. With Bose-Fermi mapping method, the exact wavefunction of quantum gas in a harmonic potential will be obtained. Based on the wavefunction, the reduced one body density matrix (ROBDM), the lowest natural orbitals and occupation numbers, and momentum distribution will be shown.

The paper is organized as follows. In Sec. II, we obtain the exact ground state wavefunction of TG gas with Bose-Fermi mapping method. In Sec. III, for different density-dependent gauge potential we present the ROBDM, momentum distributions, occupation distribution and the lowest natural orbitals. A brief summary is given in Sec. IV.

\section{Model and ground state wavefunction}
We consider Tonk-Gradeau gas consist of strongly interacting $N$ Bose atoms of mass $m$ confined in a harmonic potential
\[
V(x)=\frac12m\omega^2x^2.
\]
Following the previous procedure we introduce the density-dependent gauge potential $A(x)=g\rho(x)$ by the substitution
\begin{eqnarray}
-i\hbar \frac{d}{dx} \rightarrow -i\hbar \frac{d}{dx}-A(x).
\end{eqnarray}
Here the parameter $g$ denotes the strength of density-pendent gauge potential and $\rho(x)$ is the total atmoic density distribution of TG gas. The many body wavefunction $\Psi(x_1,x_2,\cdots,x_N)$ to describe TG gas under density-pendent gauge potential can be obtained by solving the eigen equation
\[
\hat{H}\Psi(x_1,x_2,\cdots,x_N)=E\Psi(x_1,x_2,\cdots,x_N),
\]
where the Hamiltonian should be formulated as
\begin{eqnarray}
\hat{H} = \frac{1}{2m}\sum_{i=1}^N\left[ -i\hbar \frac{d}{dx_i}-A(x_i) \right]
^{2}+V(x_i).
\end{eqnarray}
Because of the infinite strongly repulsive interaction between atoms in TG gas the many body wave function must satisfy hard core boundary condition
\begin{equation}
    \Psi(x_1,x_2,\cdots,x_N)=0 \quad \text{if} \quad  x_i=x_j.
\end{equation}

The many body wave function can be constructed with the eigenfunction $\psi_n(x)$ of the single particle Schr\"{o}dinger equation
\begin{equation}
\left[ -\frac{\hbar^2}{2m}\frac{d^2}{dx^2}+ A(x)+\frac12m\omega^2x^2\right]  \psi_n(x)=\varepsilon_n\psi_n(x).
\end{equation} 
It can be proved that
\begin{equation}
    \psi_n(x)=\phi_n(x)\exp\left[ \frac{i}{\hbar}\int^{x}A\left( x^{\prime }\right) dx^{\prime }\right],
\end{equation}
where $\phi_n(x)=\frac{e^{-x^{2}/2}H_{n}(x)}{\sqrt{\sqrt{\pi }2^{n}n!}}$ is the eigen function of one dimensional Harmonic oscillator, i.e.,
$\hat{H}_0\phi_n(x)=\varepsilon_n\phi_n(x)$ with the single particle Hamiltonian $\hat{H}_0 = -\frac{\hbar ^2}{2m} \frac{d^2}{dx^2} +\frac12m\omega^2x^2$.

Using Bose-Fermi mapping method we get the ground state wave function of TG gas under gauge potential by the polarized Fermi wavefunction $\Phi_{F}(x_{1},x_{2},\cdots ,x_{N})$
\begin{equation}
\Psi(x_{1},\cdots ,x_{N})=\mathcal{A}(x_{1},\cdots ,x_{N})\Phi
_{F}\left( x_{1},x_{2},\cdots ,x_{N}\right)
\end{equation}
with the mapping function
\begin{equation}
\mathcal{A}\left( x_{1},\cdots ,x_{N}\right) =\prod_{1\leq j<k\leq
N}sgn(x_{j}-x_{k}).
\end{equation}
The mapping function can ensure that the wave function $\Psi(x_{1},\cdots ,x_{N})$ will not change as exchanging two atoms. The wavefunction of $N$ polarized Fermion is a Slater-determinant constructed by single-particle wavefunction, i.e.,
\begin{eqnarray*}
\Phi_F(x_1,x_2,\cdots,x_N)=\frac1{\sqrt{(N)!}}\det_{j,k=1}^N\psi_j(x_k).
\end{eqnarray*}
For the case in a harmonic trap the above determinant is a Vandermonde one. Therefore using its special properties the wavefunction can be simplified as the following product form \cite{Forrester}
\begin{eqnarray}
&&\Phi _{F}\left( x_{1},x_{2},\cdots ,x_{N}\right)  \\ \nonumber
&&=\frac{1}{C_{N}^{H}}%
\prod_{j=1}^{N}\exp \left[ \frac{i}{\hbar }\int^{x_{j}}A\left( x^{\prime
}\right) dx^{\prime }\right] \\
&& \times exp(-x_{j}^{2}/2)\prod_{1\leq j<k\leq N}(x_{j}-x_{k}) \nonumber
\end{eqnarray}
with $(C_{N}^{H})^{-2}=\pi ^{-N/2}N!^{-1}\prod_{j=0}^{N-1}2^{j}j!^{-1}$.
Insert Eq. (8) into Eq. (6) we obtain the exact many body wavefunction of TG gas
\begin{eqnarray*}
\Psi \left( x_{1},x_{2},\cdots ,x_{N}\right) &=&\frac{1}{C_{N}^{H}}%
\prod_{j=1}^{N}\exp \left[ \frac{i}{\hbar }\int^{x_{j}}A\left( x^{\prime
}\right) dx^{\prime }\right] \\
&& \times exp(-x_{j}^{2}/2)\prod_{1\leq j<k\leq N}|x_{j}-x_{k}|.
\end{eqnarray*}

\section{ROBDM and momentum distribution}

Based on the above many body wavefunction, the ROBDM of TG gas can be defined as
\begin{eqnarray}
\rho (x,y)&=&N\int_{-\infty }^{\infty }dx_{1}\cdots \int_{-\infty
}^{\infty }dx_{N-1} \\ \nonumber
&&\times \Psi^{\ast }(x_{1},\cdots ,x_{N-1},x)\Psi
(x_{1},\cdots ,x_{N-1},y).
\end{eqnarray}
It is easy to prove that compared to the case without synthetic gauge potential the ROBDM of TG gas under synthetic gauge potential get a prefactor $\alpha(x,y)$ that is dependent on the gauge potential $A(x)$. That is to say, $\rho(x,y)$ can be formulated as the product of $\alpha(x,y)$ and $\rho_0(x,y)$, which is
\begin{eqnarray}
\rho \left( x,y\right) &=&\alpha(x,y)\rho_0 \left( x,y\right)
\end{eqnarray}
with
\begin{eqnarray}
\alpha \left( x,y\right) = \exp \left[ \frac{i}{\hbar }\int_{x}^{y}A\left( x^{\prime }\right)
dx^{\prime }\right].
\end{eqnarray}
$\rho _0(x,y)$ is the ROBDM of TG gas confined in a harmonic trap but without synthetic gauge potential \cite{Forrester}
\begin{eqnarray}
\rho_0 \left( x,y\right) &=&\frac{2^{N-1}}{\sqrt{\pi }\Gamma \left(
N\right) } exp(-x^{2}/2-y^{2}/2) \\  \nonumber
&&\times \det [\frac{2^{(j+k)/2}}{2\sqrt{\pi
}\sqrt{\Gamma \left( j\right) \Gamma \left( k\right) }}b_{j,k}\left(
x,y\right) ]_{j,k=1,\cdots ,N-1},
\end{eqnarray}
with
\begin{eqnarray*}
b_{j,k}\left( x,y\right) &=&\int_{-\infty }^{\infty
}dt \exp(-t^{2})|t-x||t-y|t^{j+k-2}.
\end{eqnarray*}
In the calculation $b_{j,k}(x,y)$ can be reformulated as the composition of gamma function and confluent hypergeometric function functions \cite{Forrester}.

\begin{figure}
\includegraphics[width=3.0in]{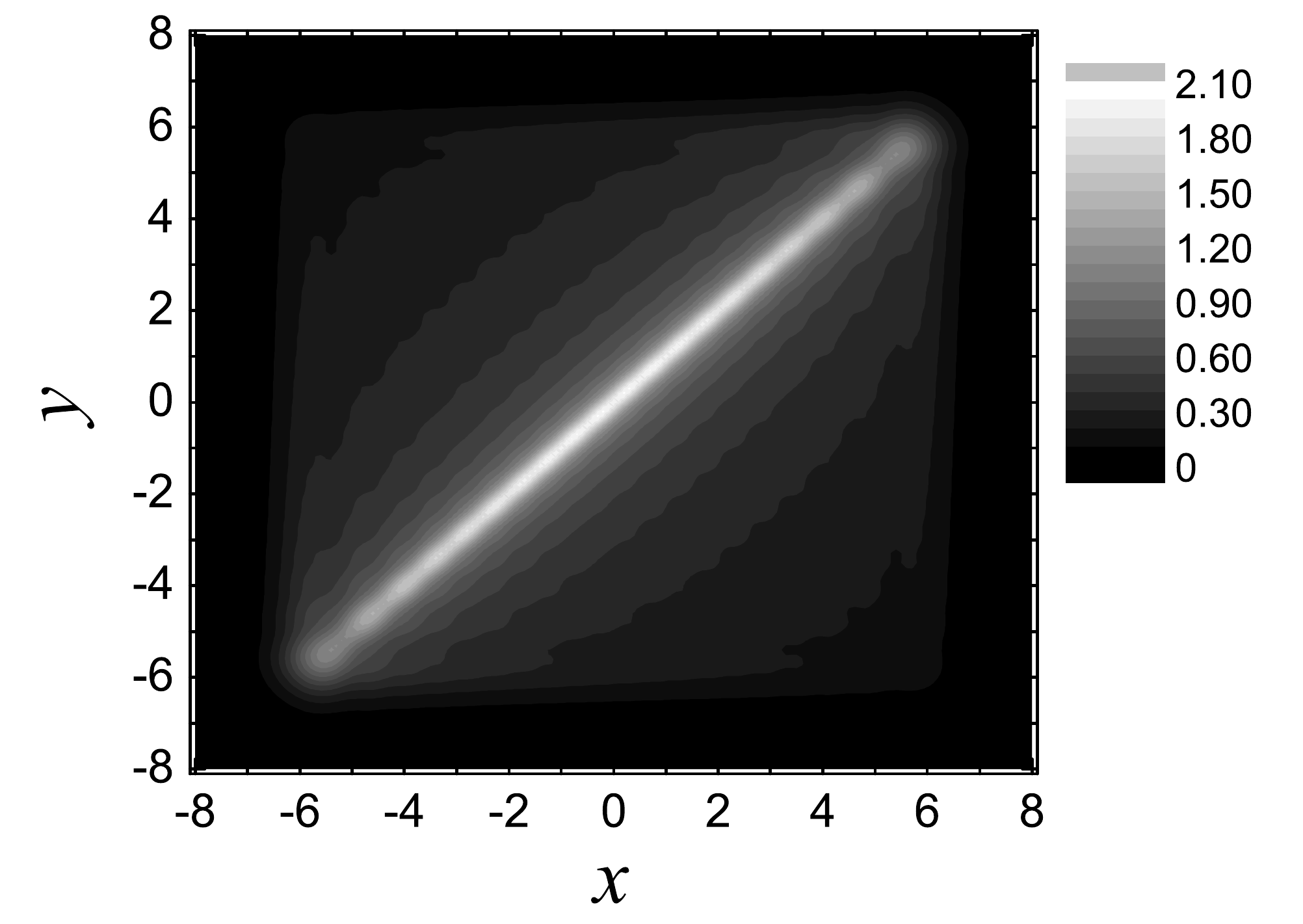}
\caption{The ROBDM $\rho_0(x,y)$ for TG gas of $N$=20 without gauge potential.}
\end{figure}
\begin{figure}
\includegraphics[width=3.5in]{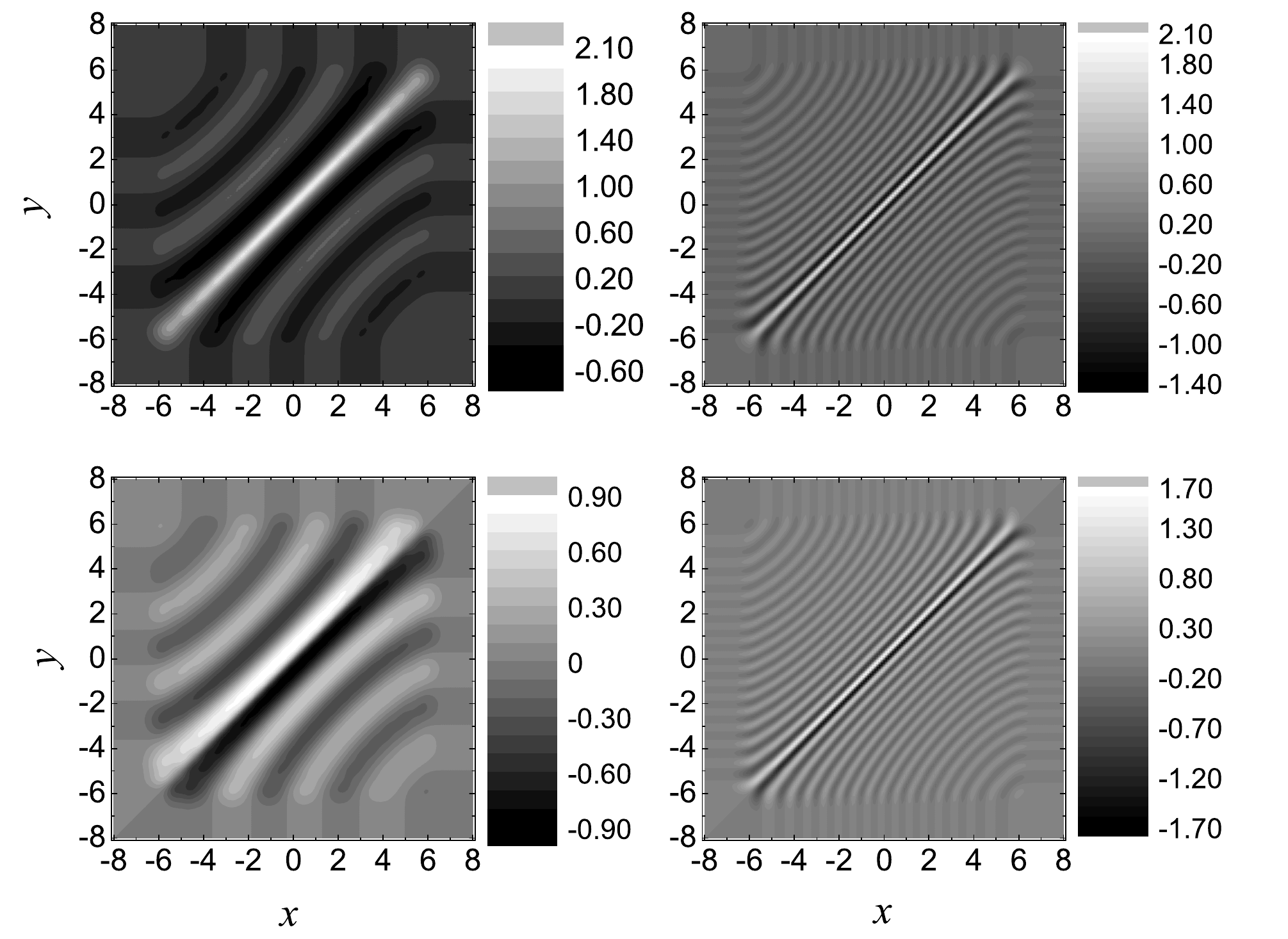}
\caption{The real part (top row) and imaginary part (below row) of ROBDM for TG gas of $N$=20 with gaute potential. Left column: g=1; Right column: g=5.}
\end{figure}

In Fig. 1 we show the ROBDM $\rho_0(x,y)$ of TG gas without gauge potential for $N$=20, while the ROBDM for the TG gas with gauge potential is displayed in Fig. 2 for the gauge potential strength $g=1$ and 5. In the former case the ROBDM is diagonal dominant and have non-negligible off-diagonal elements that is related with the off-diagonal long-range order (ODLRO) \cite{Vaidya,Girardeau}. As the synthetic gauge potential appears the ROBDM is complex rathan than real. The real part is plot in the above row and the imaginary part is plot in below row in Fig. 2. It is shown that the real part of ROBDM is also diagonal dominant and symmetric about $y=x$ and the imaginary part is asymmetric. As $|x-y|$ increases both the real part and imaginary part of ROBDM decrease in an oscillation way. That is different from the case without gauge potential in which the non-diagonal part decreases to zero monotonically with the increase of $|x-y|$. The oscillation period becomes short as the gauge potential increases.

\begin{figure}
\includegraphics[width=3.5in]{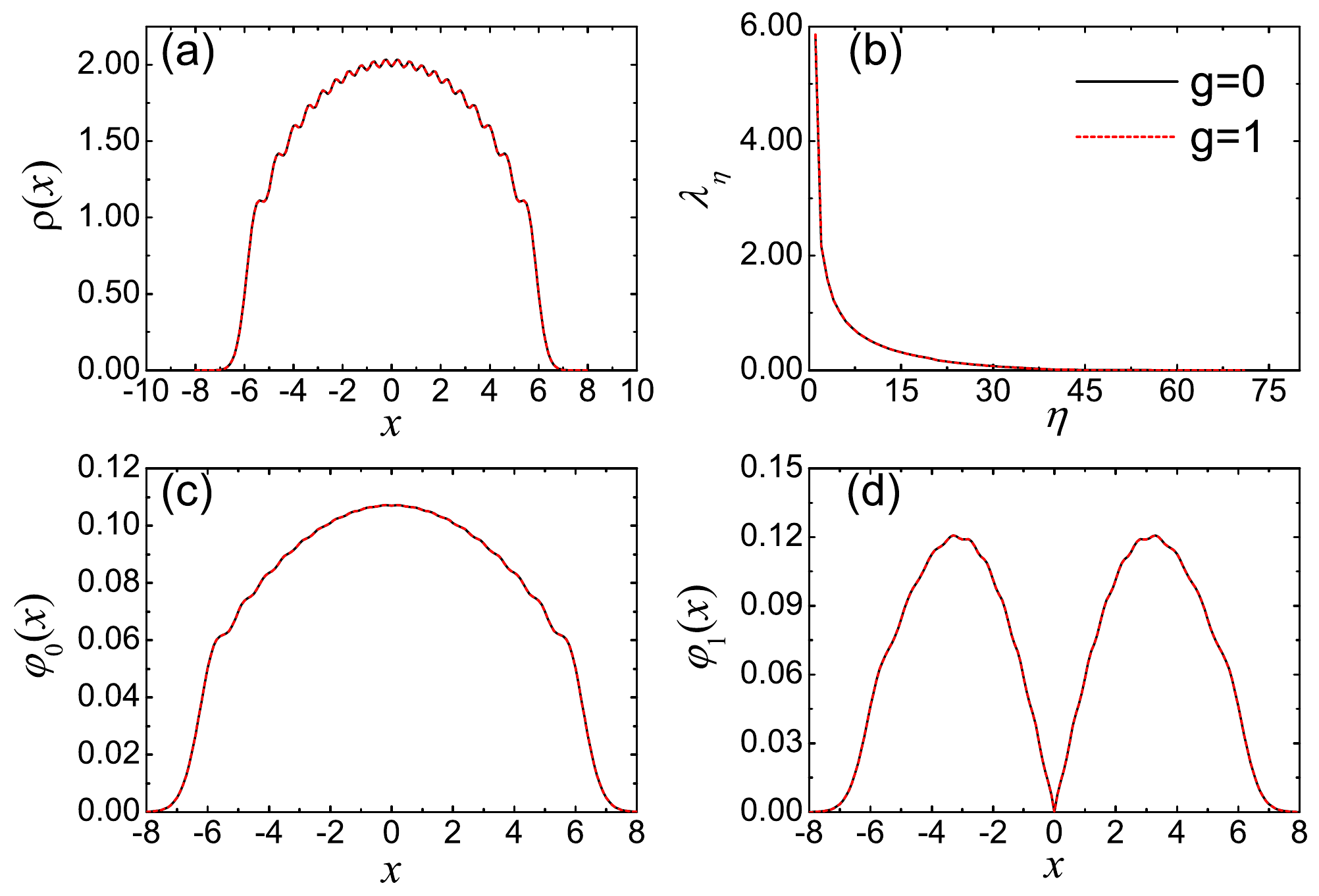}
\caption{$N$=20, $g$=0 (black solid lines) and 1 (red dashed lines). (a) The density distributions. (b) Occupation number for natural orbitals. (c) The module of the lowest natural orbital. (d) The module of the second lowest natural orbital.}
\end{figure}
According to Eq. (10) and Eq. (11) it is obvious that the density profiles of TG gas under gauge potential is irrelevant to the gauge potential, which is the diagonal part of ROBDM. We display the density distributions of TG gas with gauge potential and that without gauge potential in Fig. 3a. It is turned out that gauge potential has no effect on the density distribution of TG gas and the fermionized shell structure are exhibited.

In Fig. 3 we also plot the lowest natural orbital of TG gas $\varphi _{\eta}(x)$, which are defined as the eigenfunctions of ROBDM
\begin{eqnarray}
\int_{-\infty}^{\infty}dy\rho(x,y)\varphi  _{\eta}(y)=\lambda _{\eta}\varphi  _{\eta}(x),
\eta =1,2,....
\end{eqnarray}
Here $\lambda _{\eta}$ is the occupation number of the $\eta$th natural orbital $\varphi _{\eta}(x)$. It is deserved to notice that both the module of natural orbital and its occupation number are independent on gauge potential as shown in Fig. 3 although the ROBDM has a prefactor dependent on gauge potential. Fig. 3b show the occupation number of TG gases for $g$=0 and 1. The module of the lowest natural orbital and that of the second lowest lowest natural orbital are shown in Fig. 3c and Fig. 3d, respectively. The former has no node and there is one node in the later as if they are single particle eigen functions of ground state and the first excited state.

\begin{figure}
\includegraphics[width=3.0in]{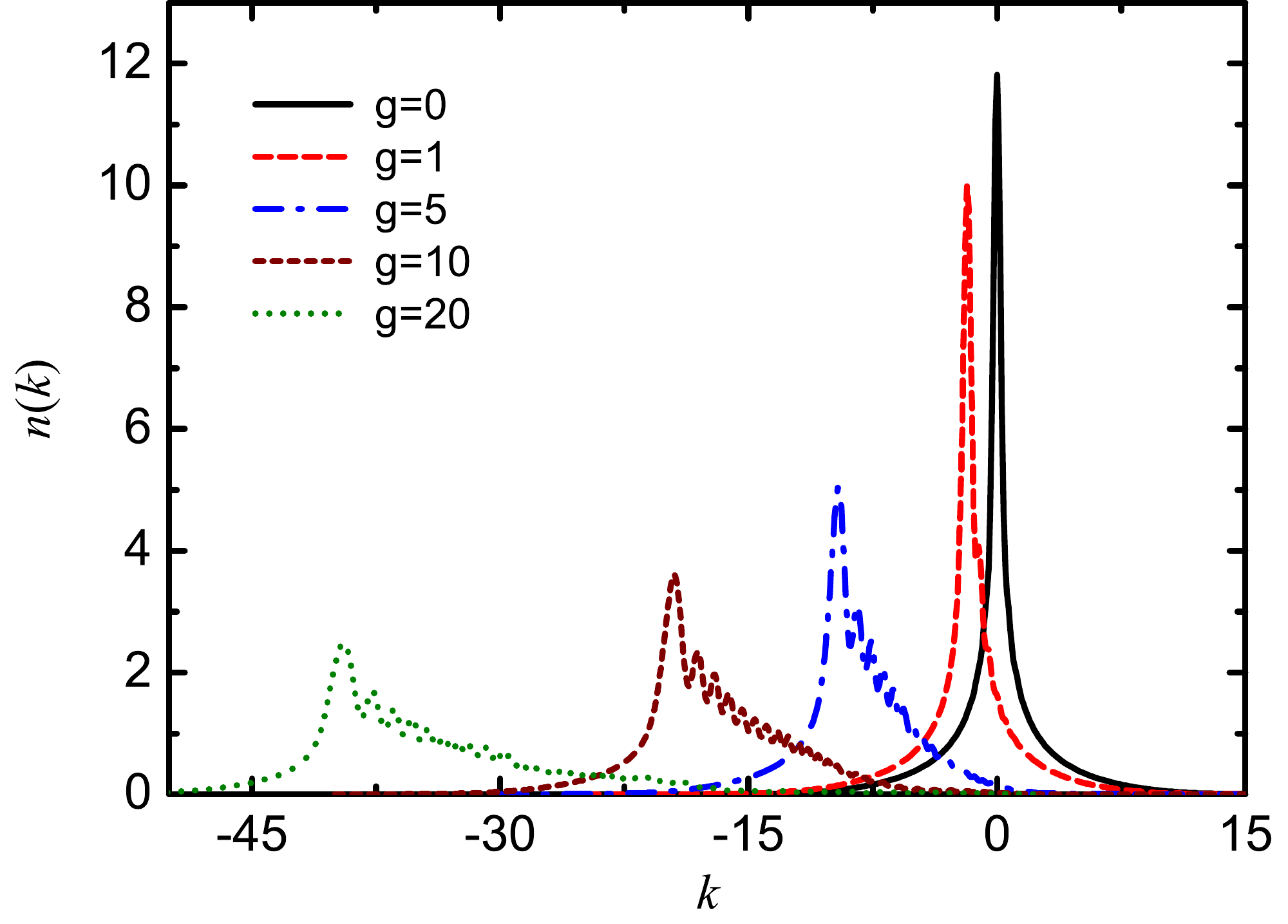}
\caption{Momentum distribution for TG gas of $N$=20 with density-dependent synthetic gauge potential $g$= 0, 1, 5, 10, 20.}
\end{figure}

Generally the momentum distribution is defined as the Fourier transformation of ROBDM
\begin{eqnarray}
n(k)=\frac{1}{2\pi}\int_{-\infty}^{\infty}dx\int_{-\infty}^{\infty}dy\rho(x,y) e^{-ik(x-y)}.
\end{eqnarray}
In the present case, we extensively define
\begin{eqnarray}
n\left( \mathbf{k}\right)  &=&\frac{1}{2\pi }\int dxdy\rho \left(
x,y\right) \exp \left[ -i\left( k_{x}x-k_{y}y\right) \right]
\end{eqnarray}
with $\mathbf{k}=(k_x,k_y)$, while Eq. (14) is the special case of $k_x=k_y=k$. Similarly, we can also obtain the Fourier transformation of $\alpha(x,y)$ and $\rho_0(x,y)$ as follows

\begin{equation*}
\alpha\left( \mathbf{k}\right)  =\frac{1}{2\pi }\int dxdy\alpha\left( x,y\right)
\exp \left[ -i\left( k_{x}x-k_{y}y\right) \right]
\end{equation*}
and
\begin{equation*}
 n_{0}\left( \mathbf{k}\right)  =\frac{1}{2\pi }\int dxdy\rho _{0}\left(
x,y\right) \exp \left[ -i\left( k_{x}x-k_{y}y\right) \right].
\end{equation*}

Inserting Eq. (10) and the inverse Fourier transformation of $\alpha(\mathbf{k})$ and $n_0(\mathbf{k})$ into Eq. (15) we get
\begin{eqnarray}
n\left( \mathbf{k}\right)
&=&\frac{1}{2\pi }\int d\mathbf{k}^{\prime }\alpha\left( \mathbf{k}^{\prime
}\right) n_{0}\left( \mathbf{k}-\mathbf{k}^{\prime }\right).
\end{eqnarray}
That is to say that the momentum distribution of the TG gas with gauge potential is the convolution of that without gauge potential to the Fourier transformation of integral of gauge potential in the region $(x,y)$, in a short word, $n\left( \mathbf{k}\right) =\frac{1}{2\pi }\alpha\left( \mathbf{k}\right) \ast n_{0}\left( \mathbf{k}\right)$.

In Fig. 4 we plot the momentum distributions of TG gases with and without synthetic gauge potential. It is shown that under the synthetic gauge potential the momentum distribution of TG gas is not symmetric any more although the momentum profiles still exhibit the typical single peak structure of Bose-like. In the situation without gauge potential Bose atoms mainly stay in the region of zero momentum and populate in the region of finite momentum in the minor probability. With the increase of the strength of gauge potential the asymmetry of momentum profiles become more obvious and the momentum profiles deviate more from the zero momentum. So under synthetic gauge potential the atoms in TG gas get finite momentum and the total momentum increases with the increase of $g$. This is on the contrary to the zero total momentum of the case without gauge potential.

\begin{figure}
\includegraphics[width=3.0in]{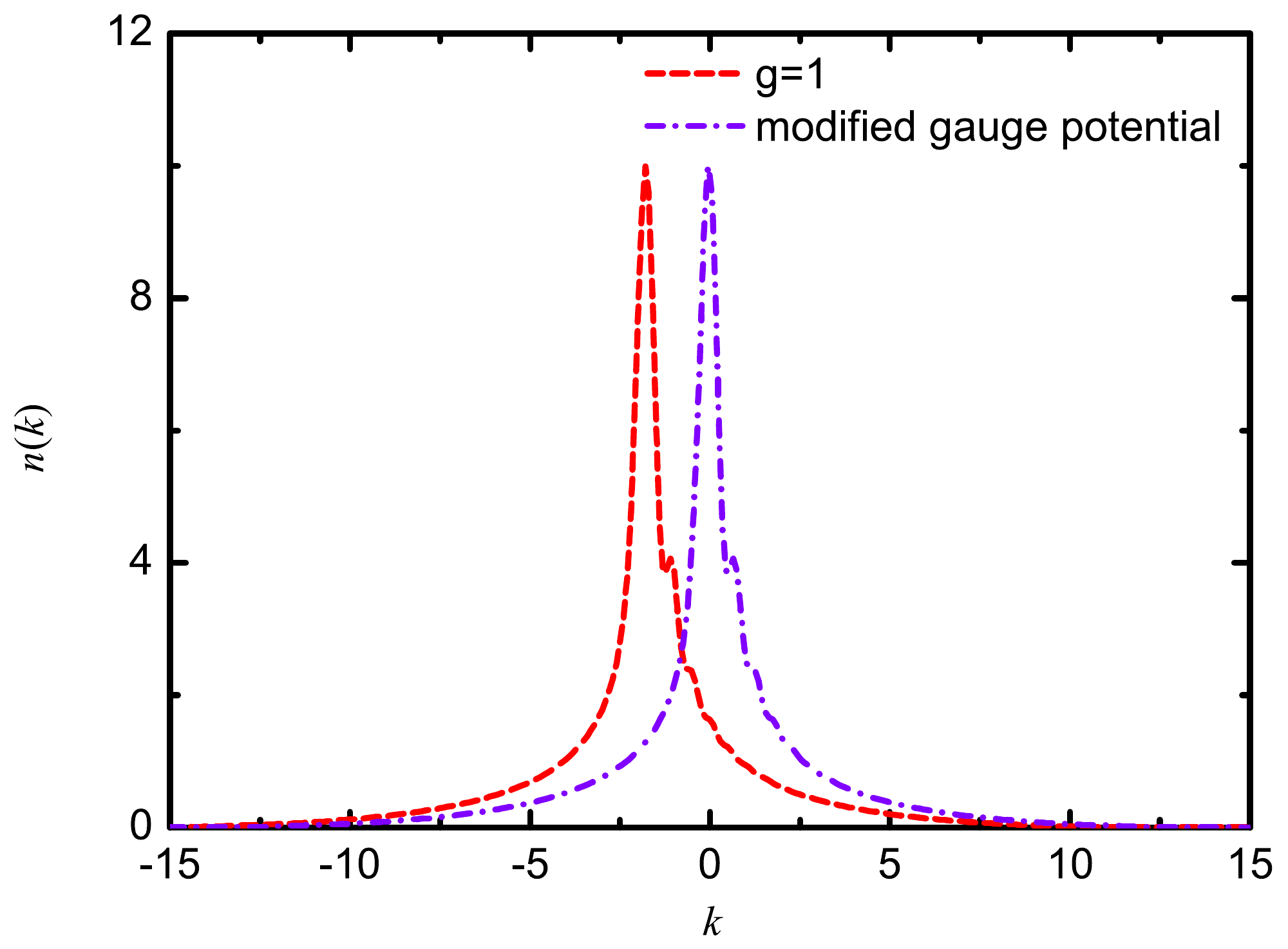}
\caption{Momentum distribution for TG gas of $N$=20 with modified density-dependent gauge potential.}
\end{figure}

The deviation of momentum profiles from zero momentum can be canceled with the replacement of modified synthetic gauge potential
\begin{eqnarray*}
A(x) = g\rho \left( x \right) +g_{0}
\end{eqnarray*}
with $g_0=\int kn(k)dk$ for $A(x)=g\rho(x)$.
It is easy to prove that in this case the momentum distribution is
\begin{eqnarray}
n\left( \mathbf{k}\right) = \frac{1}{2\pi }A\left( \mathbf{k}+\mathbf{g}_{0}\right) \ast n_{0}\left(
\mathbf{k}\right).
\end{eqnarray}
The momentum distribution for TG gas under modified gauge potential is plotted in Fig. 5. In this situation the momentum distribution preserves the exact same asymmetric profile as that of $g_0$=0 but the total momentum becomes zero.

\section{Conclusion}
In conclusion, we investigated the TG gas in a harmonic trap under synthetic gauge potential. With Bose-Fermi mapping method we obtain the exact ground state wavefunction and therefore the ROBDM, the natural orbitals and their occupation number, and the momentum distribution.

Compared with the TG gas without gauge potential, the ground state wavefunction and the ROBDM for the case with gauge potential get addition prefactors dependent on the gauge potential so that the ROBDM become complex. Its real part is diagonal dominant, which is same as the case without gauge potential. Both the real part and imaginary part of ROBDM decrease in an oscillation way rather than monotonically with the increase of $|x-y|$. It is interesting to notice that although the wavefunction is related with gauge potential, the density distribution, module of natural orbital and its occupation number are irrelevant with the synthetic gauge potential. Under the gauge potential, the momentum distribution of TG gas become asymmetry and the total momentum become finite rather than zero. The deviation from the zero momentum can be canceled by a modified synthetic gauge potential. Although the momentum distribution exhibit the similar asymmetric behaviour to that of 1D hard core anyon gas, it can be proved rigorously that the present ROBDM is not same as that of anyon gas.

\begin{acknowledgments}
This work was supported by NSF of China under Grants No. 11774026.
\end{acknowledgments}

\end{document}